\documentclass[aps,twocolumn,noshowpacs,noshowkeys,superscriptaddress
]{revtex4}
\usepackage{amsfonts}
\usepackage{amsmath}
\usepackage{amssymb}
\usepackage{graphicx}
\usepackage{textcomp}%
\begin{document}
\title{Cryogenic  properties of optomechanical silica microcavities}
\def\mpq{Max-Planck-Institut f\"ur Quantenoptik, 85748 Garching, Germany}
\def\epfl{Ecole F\'ed\'erale Polytechnique de Lausanne (EPFL),  1015 Lausanne, Switzerland}
\author{O. Arcizet}
\affiliation{\mpq}
\author{R.~Rivi\`ere}
\affiliation{\mpq}
\author{A.~Schliesser}
\affiliation{\mpq}
\author{T.~J.~Kippenberg}
\affiliation{\mpq}
\affiliation{\epfl}

\begin{abstract}
We present the optical and mechanical properties of high-Q fused silica microtoroidal resonators at cryogenic temperatures (down to 1.6\,K). A thermally induced optical multistability is observed and theoretically described; it serves to characterize quantitatively  the static heating induced by light absorption.  Moreover the  influence of structural defect states in glass on the toroid mechanical properties is observed and the resulting implications of cavity optomechanical systems on the study of mechanical dissipation discussed.
\end{abstract}
\maketitle

\textit{Introduction.---}
From their ability  to combine both high optical and mechanical properties in one and the same device, micro-toroidal optomechanical cavities can be considered as a promising  system for future progress in optomechanics at low phonon number \cite{Kippenberg2008}.
In particular these resonators support high frequency  and low mass mechanical eigen-modes  (in the range of 60\,MHz and 10\,ng, respectively) -whose clamping losses can be strongly suppressed with a suitable geometric design \cite{Anetsberger2008}- as well as simultaneously ultra-high finesse  ($10^6$) optical resonances. This feature allows to enter deeply  the resolved sideband regime \cite{Schliesser2008}, a prerequisite for cooling their radial breathing mode - a  macroscopic mechanical oscillator-  down to its quantum ground state. The cryogenic cooling of the resonator lowers the initial mean phonon occupancy of the mode of interest that can be further reduced by optical cooling \cite{Arcizet2006,Gigan2006,Schliesser2006}.
However, the specificity of the device, based on an amorphous material which confines the photons within the dielectric resonator, induces non trivial  and hereto unobserved behavior that is presented in this letter. Moreover  we discuss systems suitability for further progress towards quantum optomechanics. In particular the phonon coupling  to the structural defect states of glass opens promising perspectives for amorphous optomechanical microcavities, that may represent a new powerful  tool for probing mechanical decoherence at low temperatures\cite{Blencowe2008}.

\textit{Experimental setup.---} A tunable 1550-nm diode laser is coupled by means of a tapered fiber to an optical resonance of a microcavity, hosted in a $^4$He exchange gas cryostat. The mechanical vibrations of the toroid are imprinted on the phase of the transmitted laser field and detected via a Pound Drever Hall technique, providing a high sensitivity measurement of the toroid's displacement noise \cite{Arcizet2006a,Schliesser2008b}.
No significant modification of its opto-mechanical properties were observed when working with light intensities below $1\,\rm \mu W$.
To overcome photodetector dark noise, it is possible to take advantage of a low noise erbium doped fiber amplifier (EDFA), which when combined with a low noise Koheras fiber laser, yields a  typical displacement sensitivity  of  $3 \times\,\rm 10^{-18} m/\sqrt{Hz}$ at $60\,\rm MHz$ achieved with $1\,\rm \mu W$ of laser power injected in a  $10^5$ finesse optical resonance.
\begin{figure}[b]
\begin{center}
\includegraphics[width=\linewidth]{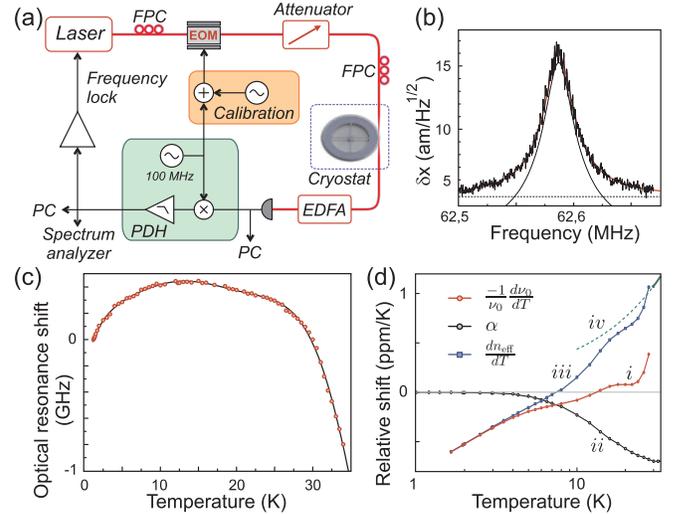}
\caption{(color online).(a): Experimental setup. PFC: polarization fiber controller, EDFA: Erbium doped fiber amplifier. (b): Typical displacement noise spectrum obtained at low temperature (1.6\,K, 4\,mbar) for the radial breathing mode of a 30-$\rm \mu m$-radius toroid oscillating at 63\,MHz. The measure of the rms amplitude  of its Brownian motion (ca. 4\,fm) serves as a toroid temperature sensor, proving its proper thermalization ensured by the exchange gas.
(c): optical resonance frequency $\nu_{o}(T)$ vs. temperature. The
solid line represents a polynomial fit of the data. (d) Relative optical frequency shift with temperature; {\it i}: $\frac{-1}{\nu_o}\frac{d\nu_o}{dT}$;  {\it ii}: silica's thermal expansion coefficient $\alpha$ \cite{White1975}; {\it ii}: inferred refractive index contribution, $\frac{dn_{\rm eff}}{dT}$; {\it iv}: measured contribution of silica's refractive, from ref.\cite{Leviton2006}, and extrapolation to lower temperatures.
}\label{fig1B}
\end{center}
\end{figure}
In this manner, the  resonator's Brownian motion  can be monitored down to 1.6\,K  (corresponding to an average occupancy of 600 phonons) with a signal-to-noise ratio ($>10\,\rm dB$ at 1.6\,K) sufficient for measuring the mechanical quality factor $Q$, the resonance frequency $ \Omega_{\rm m}/2\pi$ and the effective temperature of the radial breathing mode. The latter was compared to the cryostat temperature, determined with semiconductor sensors, and no differences (at the level of 0.5\,K) could be observed, indicating that the displacement read-out does not induce any significant temperature increase via light absorption or dynamical back action. The proper thermalization of the toroid -even for the micro-structures  weakly coupled to the silicon substrate \cite{Anetsberger2008}- requires a Helium gas pressure of ca. 5\,mbar, sufficiently low for not increasing its mechanical losses by  acoustic damping.

\textit{Temperature dependent optical resonance frequency---}
The high optical quality factor are preserved at low temperature, and no significant change (at the MHz level)  could be observed.
The frequency $\nu_o(T)$ of the microcavity optical resonance varies with the cryostat temperature (cf. Fig.\ref{fig1B}c), presenting  a $-135\,\rm MHz/K$ shift at 30\,K that reverses \cite{Park2007} around $T^\star=13.3\,\rm K$ and reaches $+100\,\rm MHz/K$ at 2\,K. The measurements were performed at low light intensity ($<100\,\rm nW$) in order to avoid inducing any thermal or optical (e.g. Kerr \cite{Treussart1998} or radiation pressure) non-linearities that could have altered the optical resonances, and at a pressure of 10\,mbar which moreover ensures a good thermalization.
The frequency shift originates  both from a mechanical expansion (thermal expansion coefficient $\alpha$)  and a change in the effective refractive index $n_{\rm eff}$ of the resonator according to the relation:  $-\frac{1}{\nu_o}\frac{d\nu_o}{dT}=\alpha +\frac{dn_{\rm eff}}{dT}$. Both silica and the surrounding medium  contribute to the effective refractive index: $n_{\rm eff}=(1-\eta)n_{\rm SiO_2}+\eta n_{\rm ext}$ where $\eta$ is the evanescent fraction of the optical mode and ranges between ca. 0.1 to $5\,\%$ depending on the spatial shape of the mode.
The inferred variation of the effective refractive index $\frac{dn_{\rm eff}}{dT}$ is shown in Fig. \ref{fig1B}d. This estimation is in agreement at high temperature (30\,K) with measurements reported in Ref. \cite{Leviton2006}.
Below 7\,K, the effective refractive index temperature dependence becomes negative. This is a consequence of the presence of the surrounding exchange gas, since additional measurements performed at 50\,mbar presented an increased negative shift, of $-520\,\rm MHz/K$ at 2\,K. From the recorded pressure evolution in the experimental chamber, the change in the Helium refractive index  have been estimated \cite{Luther1996} at a level of ca.$-10\,\rm ppm/K$ at 5\,K, which is in agreement with the measured value of ($-.15\,\rm ppm/K$).  A quantitative study of the surrounding Helium contribution will allow to measure for the first time silica's optical refractive index at low temperatures (note that the temperature dependence of silica's microwave dielectric constant is known to reverse \cite{Schickfus1976}).
The inversion  takes on high practical interest since it renders the cooling side of the resonance thermally stable \cite{Carmon2004}, as opposed to room temperature situation.

\begin{figure}[h]
\begin{center}
\includegraphics[width=.95 \linewidth]{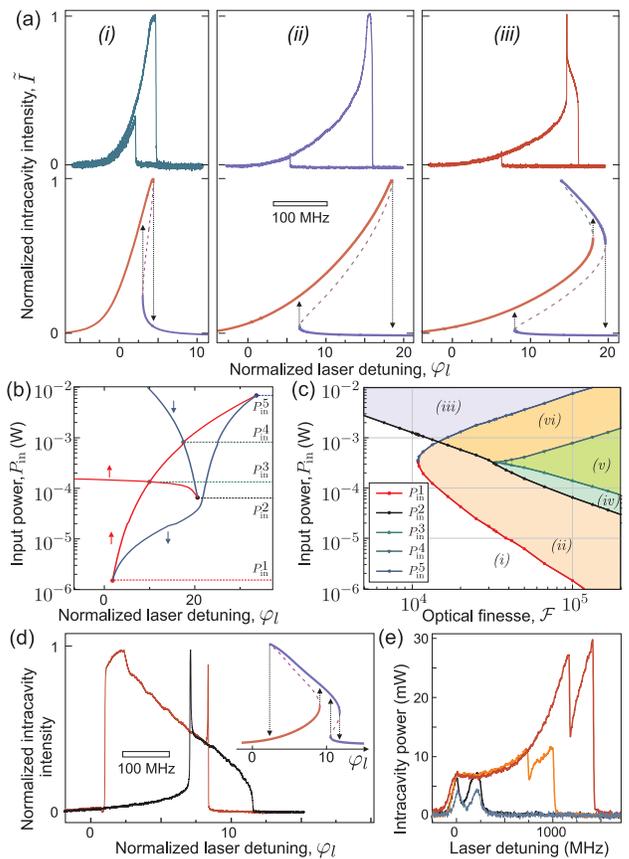}
\end{center}
\caption{(color online). (a) Above:  normalized intracavity intensity $\tilde I$  versus laser normalized detuning  $\varphi_l$ for increasing input powers  (resp. $13$, $131$ and $260\,\rm \mu W$) presenting a multistable behavior (finesse $\mathcal{F} = 44\,000$, optical linewidth $ 2 \Omega_{\rm c}/2\pi= \,\rm 28\, MHz $ critically coupled, $T_0=2.3\,\rm K$). Below: results of the simulations for increasing   $\mu$ parameter (resp. 0.8, 8.4 and 16.7).
(b) laser detuning of the four turning points as a function of the input power ($10^5$ finesse,  4\,K and  $R=30\,\rm \mu m$).  Their relative position defines the different observable regimes, that are reported in panel (c) as a function of the cavity finesse. (c):($\it i$): monostability; ($\it ii$): "red side" bistability; ($\it iii$): ``blue side" bistability; ($\it iv,v$): tristability; ($\it vi$) double bistability.
(d) Typical transmission curve obtained for a lower exchange gas pressure ($0.5\,\rm mbar$, $P_{\rm in}= 280\,\rm\mu W$, $\mathcal{F}=31\,000$) presenting a double bistable behavior (region $\it vi$).
(e) Typical (splitted) cavity resonances  observed for higher He pressure ($70\,\rm mbar$, $T=2.0\,\rm K$, $\mathcal{F}=17\,000$) for increasing input power (resp. 3, 5, 8 and $16\,\rm \mu W$) that reveal the formation of a superfluid Helium film on the microcavity (see text). }\label{fig3}
\end{figure}

\textit{Optical multistability.---}
The measured temperature dependence  is responsible for a thermally induced multistability. A fraction of the light is absorbed inside the silica resonator whose temperature increases and induces an optical frequency shift that distorts the Lorentzian profile of the resonances. If the optical resonance is shifted by more than half its linewidth, a bistability appears, as can be observed in Fig.\,\ref{fig3}a ($\it i$).
It can only be observed for resonances whose half linewidth is smaller than the optical frequency shift between the cryostat temperature $T_0$  and $T^\star$ (i.e. $\mathcal{F} > 10\,000$ at 4\,K for a radius of $30 \,\rm\mu m$).
For increasing input intensities, it is possible to observe  a regime  where the cavity transmission presents two turning points. This is a consequence of the inversion of the optical frequency shift that creates additional working points if the light induced temperature increase is strong enough to reach the inversion temperature ($T^\star$).

To model the multi-stability, one has to take into account the coupling between the optical field $\alpha$ and the effective resonator temperature $T$. The normalized complex intracavity optical field $\tilde\alpha\equiv \alpha/\sqrt{I_{\rm max}}$,  where $I_{\rm max}\equiv I_{\rm in} K \frac{\mathcal{F}}{\pi}$ is the maximum intracavity photon flux expected for a  input flux $I_{\rm in}$ and coupling efficiency $K$ ($K=1$ in case of impedance matching), follows the dynamical equation:
$
\frac{1}{\Omega_{\rm c}}\frac{d\tilde\alpha}{dt}= \left(-1+i \varphi(t,T)\right)\tilde\alpha + 1$,
where the laser to cavity detuning $\varphi$ is normalized to the cavity bandwidth $\Omega_{\rm c}/2\pi=c/(4\pi  n R \mathcal{F})$ and can be written as $\varphi(t,T)=\varphi_l(t)-\frac{4\pi n R\mathcal{F}}{c}\nu_o(T_0+\delta T(t))$ where $\varphi_l$ stands for the laser frequency scan and $\delta T$ is the effective temperature increase induced by light absorption. The dynamical equivalent temperature response depends on the mechanism considered (expansion or refraction), we define here a global thermal response function  $\chi_{\rm th}$ for the effective temperature increase, $\delta T (t)= h \nu I_{\rm max} \int_{-\infty}^t{\chi_{\rm th}(t-t')\tilde I (t') dt'}$, where  $\tilde I= |\alpha|^2 /I_{\rm max}$  is the normalized intracavity intensity.
The static working point of the system is then solution of  $1+ (\varphi_l-\frac{4\pi n R \mathcal{F}}{c}\nu_o(T_0+ \mu \tilde I))^2= 1/{\tilde I}$,
with $\mu\equiv  h\nu\bar I_{\rm max} \bar\chi_{\rm th}^{\rm stat} $ being the maximum light induced heating expected, $\bar\chi_{\rm th}^{\rm stat}=\int_{-\infty}^\infty{ \chi_{\rm th}(t)dt}$, in $\rm K/W$  representing the  induced static heating per intracavity power.
A fit (with a $7^{\rm th}$ order polynome) of the experimental $\nu_o(T)$ is used for the numerical resolution of the non-linear equation. It allows to calculate the normalized intracavity intensity for each laser detuning $\varphi_l$ and for various input intensities $I_{\rm in}$ by varying the parameter $\mu$.
The results of the simulation are shown in Fig.\,\ref{fig3}a and present a remarkable agreement with experimental traces. The dynamically unstable branches are marked with dashed lines. Compared to higher temperature behavior, the upper stable branch does not present the typical triangular shape \cite{Carmon2004} (also observed for Kerr or radiation pressure non-linearities) but is progressively curved as the temperature approaches $T^\star$ ({\it i}, {\it ii}).
When the input intensity is sufficient for heating the resonator beyond $T^\star$, two new branches appear, allowing the onset of a novel multistability ({\it iii}). At even higher intensities, the upper branch can be found at a laser detuning lower than the "blue side" lower turning point (cf. Fig.\,\ref{fig3}d inset), inducing a characteristic double turning point feature when reducing the laser frequency.
This mechanism also allows to quantify the temperature increase induced by light absorption (i.e. the parameter $\bar\chi_{\rm th}^{\rm stat}$). A simple estimation can be obtained by measuring the injected power $P_{\rm in}^{\rm thres}$ required for observing the multistability (or $\mu = T^\star-T_0$) and the cavity optical parameters (coupling and internal losses ). A value of $\bar\chi_{\rm th}^{\rm stat} \approx 4.5\,\rm K/W $ is extracted for an exchange gas pressure of $5\,\rm mbar$. For lower pressures of exchange gas, it increases up to $8.6\,\rm K/W$ at $0.5\,\rm mbar$ (assuming the same static optical shift), emphasizing its crucial role for thermalization.
The static heating of the resonator is of particular relevance in the context of ground state cooling  of mechanical modes that requires high optical cooling power (up to ca. $1\,\rm mW$), since it competes with the laser cooling. It can however be strongly circumvented  in  the  resolved sideband regime where the laser is far detuned from the optical resonance.

For completeness, it is important to mention that at higher exchange gas pressure, it is possible to enter the $^4$He superfluid phase where a thin layer of Helium appears on the surface of the microcavity. A clear signature of this phenomena is imprinted on the microcavity's optical transmission (cf. Fig. \ref{fig3}e).  At low light intensities, the cavity resonances are undeformed and simultaneously  slightly red  shifted (ca. -40\,MHz). It corresponds to the build-up in presence of  the superfluid layer whose refractive index temperature dependence reverses below the lambda point \cite{Edwards1956} and underlines the increased heat extraction efficiency. If the intracavity intensity exceeds 6.3\,mW,  the optical resonances are seriously deformed, presenting a blue shift of the maximum build-up that  is found again to strongly depend on the injected power.
Note also that the total optical frequency shift observed  ($>1\,\rm GHz$) at 70\,mbar is higher than the 350 \,MHz measured at 10\,mbar,  underlining the  role  of the surrounding Helium in the static frequency shift observed. Microcavities constitute a promising tool for non destructive studies of Helium films, and will be subject of further report.

\textit{Mechanical properties ---} The amorphous nature of the micro-resonator strongly influences its mechanical behavior.
Indeed,  the experiment provides a novel way of probing phonon propagation properties (speed of sound and attenuation) in glass at low temperatures \cite{Blencowe2008}, that can be deduced from the micro-resonator's Brownian motion (respectively its mechanical frequency and  quality factor).
\begin{figure}[t]
\begin{center}
\includegraphics[width=\linewidth]{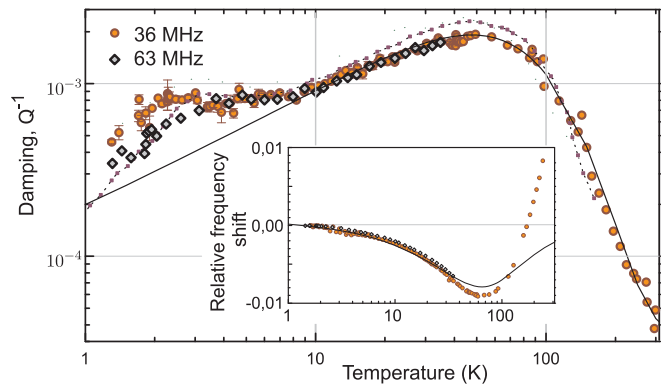}
\caption{ Mechanical damping $Q^{-1}$, and relative frequency shift $\delta\Omega_{\rm m}/\Omega_{\rm m}$, versus temperature for 36-MHz- (red circles) and 63-MHz- (black squares) resonators.  The temperature dependence observed reveal the coupling between phonon and structural defects of glass, modelized by an assembly of TLS. The solid line is a fit of the thermally activated regime (above 10\,K) according to \cite{Vacher2005}. The dotted line corresponds to an acoustic wave attenuation experiment at 40\,MHz \cite{Bartell1982}.  }\label{fig4}
\end{center}
\end{figure}
Results are represented in Fig.\,\ref{fig4} and present a non trivial temperature dependence that is characteristic of amorphous materials.
Indeed, most of their acoustic and dielectric properties, ranging from kHz to GHz frequencies,  that have been widely studied in bulk materials, can be explained \cite{Enss2005,Jackle1972} by considering a coupling of the strain and electromagnetic fields to the structural defects of glass -whose origin is still under investigation- modelized as an assembly of two-level systems (TLS)  presenting a wide distribution of energy parameters.
Interestingly, our approach consisting of a high sensitivity detection of the thermal displacement noise in a micrometric acoustic resonator -the least perturbing measurement- gives results in good agreement with direct studies of sound propagation in bulk material, relying on an external acoustic driving.

The mechanical damping presents a maximum at ca. 50\,K  of $Q\approx 500$, corresponding to the thermally activated relaxation regime. It is fitted here with the parameters taken from \cite{Vacher2005}. Below 10\,K the relaxation mechanism is dominated by tunneling assisted transitions between the two levels of the defects. They are responsible for both the plateau ($Q\approx 1200$ at 5\,K) and the further improvement of the mechanical quality factor observed at lower temperatures and higher frequencies ($Q\propto \Omega_{\rm m}/ T^{3}$).

In the context of ground state cooling, it is important to maintain a high mechanical quality factor at low temperatures for facilitating the readout and providing an efficient temperature reduction with optical cooling techniques. In this view, a promising  mechanical Q factor above 30\,000 for 90\,MHz oscillators can be expected at 600\,mK \cite{Enss2005}, a temperature at which the $\rm ^3He$ vapor pressure is still high enough  (ca. 1\,mbar) to ensure a proper thermalization of the resonator in the resolved sideband regime.

It is important to mention here that the phonon coupling to the assembly of two-level systems opens promising future perspectives for amorphous materials based optomechanical devices. Indeed it has been shown that in addition to the relaxation mechanisms described above there also exists  processes  where the phonon resonantly  interacts with the  TLS presenting the right energy splitting \cite{Jackle1972}. This mechanism dominates the phonon propagation properties at lower temperatures and higher frequencies (and is responsible for the anomalous electromagnetic dispersion previously mentioned). At 1\,K and 500\,MHz, their contributions have similar magnitude \cite{Jackle1972,Enss2005}. Our system has the potential for entering this resonant regime since higher frequency oscillators can be obtained by reducing the toroid size and working on higher order radial breathing modes, that could be still thermalized and monitored in $^3$He cryostats.

There exists a finite density of TLS and it has been shown that they can be saturated at both high electromagnetic or acoustic intensities \cite{Hunklinger1972} ($J_{\rm ac}^{\rm sat}\approx 10^{-3}\,\rm W/m^2$ at 1\,K). The latter would require an mechanical driving of $\delta x = \sqrt{ \frac{2 J_{\rm ac}^{\rm sat}}{\rho c_s\Omega_{\rm m}^2}}\approx 2\,\rm fm$ that is experimentally feasible via radiation pressure force and well within the detection capacity offered by optical readout.
Importantly, the TLS can be  coupled simultaneously to strain and electromagnetic fields and cross-couplings have been demonstrated \cite{Laermans1977} at microwave intensities ($20\,\rm W/m^2$ at 1\,K) that could feasibly be implemented. This would allow a possible control and read-out of the resonator mechanical state by mean of an external radio frequency field, a promising step towards mechanical quantum state engineering. Furthermore, this interaction has allowed to generate phonon echo phenomena  \cite{Golding1976} at temperatures and frequencies where the equivalent mechanical oscillator would be found at occupancies close to unity and could be interestingly exploited to measure mechanical decoherence at low phonon number \cite{Armour2008}.

This work was supported by an Independent Max Planck Junior Research Group of the Max Planck Society, the Deutsche Forschungsgemeinschaft (DFG-GSC) and a Marie Curie Excellence Grant. O.A. acknowledges funding from a Marie Curie Grant (QUOM).


\begin{thebibliography}{25}
\expandafter\ifx\csname natexlab\endcsname\relax\def\natexlab#1{#1}\fi
\expandafter\ifx\csname bibnamefont\endcsname\relax
  \def\bibnamefont#1{#1}\fi
\expandafter\ifx\csname bibfnamefont\endcsname\relax
  \def\bibfnamefont#1{#1}\fi
\expandafter\ifx\csname citenamefont\endcsname\relax
  \def\citenamefont#1{#1}\fi
\expandafter\ifx\csname url\endcsname\relax
  \def\url#1{\texttt{#1}}\fi
\expandafter\ifx\csname urlprefix\endcsname\relax\def\urlprefix{URL }\fi
\providecommand{\bibinfo}[2]{#2}
\providecommand{\eprint}[2][]{\url{#2}}

\bibitem[{\citenamefont{Kippenberg et~al.}(2008)\citenamefont{Kippenberg and Vahala}}]{Kippenberg2008}
  \bibinfo{author}{\bibfnamefont{T.~J.} \bibnamefont{Kippenberg}} \bibnamefont{and}
   \bibinfo{author}{\bibfnamefont{K.~J.} \bibnamefont{Vahala}},
  \bibinfo{journal}{Science} \textbf{\bibinfo{volume}{321}},
  \bibinfo{pages}{1172} (\bibinfo{year}{2008}).




\bibitem[{\citenamefont{Anetsberger et~al.}(2008)\citenamefont{Anetsberger,
  Rivi\`{e}re, Schliesser, Arcizet, and Kippenberg}}]{Anetsberger2008}
\bibinfo{author}{\bibfnamefont{G.}~\bibnamefont{Anetsberger}},
  \bibinfo{author}{\bibfnamefont{R.}~\bibnamefont{Rivi\`{e}re}},
  \bibinfo{author}{\bibfnamefont{A.}~\bibnamefont{Schliesser}},
  \bibinfo{author}{\bibfnamefont{O.}~\bibnamefont{Arcizet}}, \bibnamefont{and}
  \bibinfo{author}{\bibfnamefont{T.~J.} \bibnamefont{Kippenberg}},
  \bibinfo{journal}{Nature Photon.} \textbf{\bibinfo{volume}{2}},
  \bibinfo{pages}{627} (\bibinfo{year}{2008}).

\bibitem[{\citenamefont{Schliesser
  et~al.}(2008{\natexlab{a}})\citenamefont{Schliesser, Rivi\`{e}re,
  Anetsberger, Arcizet, and Kippenberg}}]{Schliesser2008}
\bibinfo{author}{\bibfnamefont{A.}~\bibnamefont{Schliesser}},
  \bibinfo{author}{\bibfnamefont{R.}~\bibnamefont{Rivi\`{e}re}},
  \bibinfo{author}{\bibfnamefont{G.}~\bibnamefont{Anetsberger}},
  \bibinfo{author}{\bibfnamefont{O.}~\bibnamefont{Arcizet}}, \bibnamefont{and}
  \bibinfo{author}{\bibfnamefont{T.~J.} \bibnamefont{Kippenberg}},
  \bibinfo{journal}{Nat. Phys.} \textbf{\bibinfo{volume}{4}},
  \bibinfo{pages}{415} (\bibinfo{year}{2008}{\natexlab{a}}).

\bibitem[{\citenamefont{Arcizet
  et~al.}(2006{\natexlab{a}})\citenamefont{Arcizet, Cohadon, Briant, Pinard,
  and Heidmann}}]{Arcizet2006}
\bibinfo{author}{\bibfnamefont{O.}~\bibnamefont{Arcizet}},
  \bibinfo{author}{\bibfnamefont{P.-F.} \bibnamefont{Cohadon}},
  \bibinfo{author}{\bibfnamefont{T.}~\bibnamefont{Briant}},
  \bibinfo{author}{\bibfnamefont{M.}~\bibnamefont{Pinard}}, \bibnamefont{and}
  \bibinfo{author}{\bibfnamefont{A.}~\bibnamefont{Heidmann}},
  \bibinfo{journal}{Nature} \textbf{\bibinfo{volume}{444}}, \bibinfo{pages}{71}
  (\bibinfo{year}{2006}{\natexlab{a}}).

\bibitem[{\citenamefont{Gigan et~al.}(2006)\citenamefont{Gigan, Bohm,
  Paternostro, Blaser, Langer, Hertzberg, Schwab, Bauerle, Aspelmeyer, and
  Zeilinger}}]{Gigan2006}
\bibinfo{author}{\bibfnamefont{S.}~\bibnamefont{Gigan}}
  \bibnamefont{{\emph et al}},
  \bibinfo{journal}{Nature} \textbf{\bibinfo{volume}{444}}, \bibinfo{pages}{67}
  (\bibinfo{year}{2006}).


\bibitem[{\citenamefont{Schliesser et~al.}(2006)\citenamefont{Schliesser,
  Del'Haye, Nooshi, Vahala, and Kippenberg}}]{Schliesser2006}
\bibinfo{author}{\bibfnamefont{A.}~\bibnamefont{Schliesser}},
  \bibinfo{author}{\bibfnamefont{P.}~\bibnamefont{Del'Haye}},
  \bibinfo{author}{\bibfnamefont{N.}~\bibnamefont{Nooshi}},
  \bibinfo{author}{\bibfnamefont{K.~J.} \bibnamefont{Vahala}},
  \bibnamefont{and} \bibinfo{author}{\bibfnamefont{T.~J.}
  \bibnamefont{Kippenberg}}, \bibinfo{journal}{Phys. Rev. Lett.}
  \textbf{\bibinfo{volume}{97}}, \bibinfo{pages}{243905}
  (\bibinfo{year}{2006}).

\bibitem[{\citenamefont{Blencowe}(2008)}]{Blencowe2008}
\bibinfo{author}{\bibfnamefont{M.~P.} \bibnamefont{Blencowe}},
  \bibinfo{journal}{Nature Physics} \textbf{\bibinfo{volume}{4}},
  \bibinfo{pages}{753} (\bibinfo{year}{2008}).

\bibitem[{\citenamefont{Arcizet
  et~al.}(2006{\natexlab{b}})\citenamefont{Arcizet, Cohadon, Briant, Pinard,
  Heidmann, Mackowski, Michel, Pinard, Fran\c{c}ais, and
  Rousseau}}]{Arcizet2006a}
\bibinfo{author}{\bibfnamefont{O.}~\bibnamefont{Arcizet}}
    \bibnamefont{{\emph et al}},
  \bibinfo{journal}{Phys. Rev. Lett.} \textbf{\bibinfo{volume}{97}},
  \bibinfo{eid}{133601} (\bibinfo{year}{2006}{\natexlab{b}}).

\bibitem[{\citenamefont{Schliesser
  et~al.}(2008{\natexlab{b}})\citenamefont{Schliesser, Anetsberger, Rivi\`ere,
  Arcizet, and Kippenberg}}]{Schliesser2008b}
\bibinfo{author}{\bibfnamefont{A.}~\bibnamefont{Schliesser}},
  \bibinfo{author}{\bibfnamefont{G.}~\bibnamefont{Anetsberger}},
  \bibinfo{author}{\bibfnamefont{R.}~\bibnamefont{Rivi\`ere}},
  \bibinfo{author}{\bibfnamefont{O.}~\bibnamefont{Arcizet}}, \bibnamefont{and}
  \bibinfo{author}{\bibfnamefont{T.~J.} \bibnamefont{Kippenberg}},
  \bibinfo{journal}{New J. Phys.} \textbf{\bibinfo{volume}{10}},
  \bibinfo{pages}{095015} (\bibinfo{year}{2008}{\natexlab{b}}).

\bibitem[{\citenamefont{White}(1975)}]{White1975}
\bibinfo{author}{\bibfnamefont{G.~K.} \bibnamefont{White}},
  \bibinfo{journal}{Phys. Rev. Lett.} \textbf{\bibinfo{volume}{34}},
  \bibinfo{pages}{204} (\bibinfo{year}{1975}).

\bibitem[{\citenamefont{Leviton and Frey}(2006)}]{Leviton2006}
\bibinfo{author}{\bibfnamefont{D.~B.} \bibnamefont{Leviton}} \bibnamefont{and}
  \bibinfo{author}{\bibfnamefont{B.~J.} \bibnamefont{Frey}},
  \bibinfo{journal}{Proc. SPIE} \textbf{\bibinfo{volume}{6273}},
  \bibinfo{pages}{62732K} (\bibinfo{year}{2006}).

\bibitem[{\citenamefont{Park and Wang}(2007)}]{Park2007}
\bibinfo{author}{\bibfnamefont{Y.-S.} \bibnamefont{Park}} \bibnamefont{and}
  \bibinfo{author}{\bibfnamefont{H.}~\bibnamefont{Wang}},
  \bibinfo{journal}{Opt. Lett.} \textbf{\bibinfo{volume}{32}},
  \bibinfo{pages}{3104} (\bibinfo{year}{2007}).

\bibitem[{\citenamefont{Treussart et~al.}(1998)\citenamefont{Treussart,
  Ilchenko, Roch, Hare, Lef\`{e}vre-Seguin, Raimond, and
  Haroche}}]{Treussart1998}
\bibinfo{author}{\bibfnamefont{F.}~\bibnamefont{Treussart}}
   \bibnamefont{{\emph et al}},
  \bibinfo{journal}{Eur. Phys. J. D} \textbf{\bibinfo{volume}{1}},
  \bibinfo{pages}{235} (\bibinfo{year}{1998}).

\bibitem[{\citenamefont{Luther et~al.}(1996)\citenamefont{Luther, Grohmann, and
  Fellmuth}}]{Luther1996}
\bibinfo{author}{\bibfnamefont{H.}~\bibnamefont{Luther}},
  \bibinfo{author}{\bibfnamefont{K.}~\bibnamefont{Grohmann}}, \bibnamefont{and}
  \bibinfo{author}{\bibfnamefont{B.}~\bibnamefont{Fellmuth}},
  \bibinfo{journal}{Metrologia} \textbf{\bibinfo{volume}{33}},
  \bibinfo{pages}{341 } (\bibinfo{year}{1996}).

\bibitem[{\citenamefont{von Schickfus and Hunklinger}(1976)}]{Schickfus1976}
\bibinfo{author}{\bibfnamefont{M.}~\bibnamefont{von Schickfus}}
  \bibnamefont{and}
  \bibinfo{author}{\bibfnamefont{S.}~\bibnamefont{Hunklinger}},
  \bibinfo{journal}{J. Phys. C} \textbf{\bibinfo{volume}{9}},
  \bibinfo{pages}{L439} (\bibinfo{year}{1976}).

\bibitem[{\citenamefont{Carmon et~al.}(2004)\citenamefont{Carmon, Yang, and
  Vahala}}]{Carmon2004}
\bibinfo{author}{\bibfnamefont{T.}~\bibnamefont{Carmon}},
  \bibinfo{author}{\bibfnamefont{L.}~\bibnamefont{Yang}}, \bibnamefont{and}
  \bibinfo{author}{\bibfnamefont{K.}~\bibnamefont{Vahala}},
  \bibinfo{journal}{Opt. Express} \textbf{\bibinfo{volume}{12}},
  \bibinfo{pages}{4742} (\bibinfo{year}{2004}).

\bibitem[{\citenamefont{Edwards}(1956)}]{Edwards1956}
\bibinfo{author}{\bibfnamefont{M.~H.} \bibnamefont{Edwards}},
  \bibinfo{journal}{Can. J. Phys.} \textbf{\bibinfo{volume}{34}},
  \bibinfo{pages}{898} (\bibinfo{year}{1956}).

\bibitem[{\citenamefont{Vacher et~al.}(2005)\citenamefont{Vacher, Courtens, and
  Foret}}]{Vacher2005}
\bibinfo{author}{\bibfnamefont{R.}~\bibnamefont{Vacher}},
  \bibinfo{author}{\bibfnamefont{E.}~\bibnamefont{Courtens}}, \bibnamefont{and}
  \bibinfo{author}{\bibfnamefont{M.}~\bibnamefont{Foret}},
  \bibinfo{journal}{Phys. Rev. B} \textbf{\bibinfo{volume}{72}},
  \bibinfo{eid}{214205} (\bibinfo{year}{2005}).

\bibitem[{\citenamefont{Bartell and Hunklinger}(1982)}]{Bartell1982}
\bibinfo{author}{\bibfnamefont{U.}~\bibnamefont{Bartell}} \bibnamefont{and}
  \bibinfo{author}{\bibfnamefont{S.}~\bibnamefont{Hunklinger}},
  \bibinfo{journal}{J. Phys. (Paris) Colloq.} \textbf{\bibinfo{volume}{43}},
  \bibinfo{pages}{C9–} (\bibinfo{year}{1982}).

\bibitem[{\citenamefont{Enss and Hunklinger}(2005)}]{Enss2005}
\bibinfo{author}{\bibfnamefont{C.}~\bibnamefont{Enss}} \bibnamefont{and}
  \bibinfo{author}{\bibfnamefont{S.}~\bibnamefont{Hunklinger}},
  \emph{\bibinfo{title}{Low-Temperature Physics}}
  (\bibinfo{publisher}{Springer}, \bibinfo{place}{Heidelberg}, \bibinfo{year}{2005}).

\bibitem[{\citenamefont{J\"{a}ckle}(1972)}]{Jackle1972}
\bibinfo{author}{\bibfnamefont{J.}~\bibnamefont{J\"{a}ckle}},
  \bibinfo{journal}{Z. Phys.} \textbf{\bibinfo{volume}{257}},
  \bibinfo{pages}{212} (\bibinfo{year}{1972}).

\bibitem[{\citenamefont{Hunklinger et~al.}(1972)\citenamefont{Hunklinger,
  Arnold, Stein, Nava, and Dransfeld}}]{Hunklinger1972}
\bibinfo{author}{\bibfnamefont{S.}~\bibnamefont{Hunklinger}},
  \bibinfo{author}{\bibfnamefont{W.}~\bibnamefont{Arnold}},
  \bibinfo{author}{\bibfnamefont{S.}~\bibnamefont{Stein}},
  \bibinfo{author}{\bibfnamefont{R.}~\bibnamefont{Nava}}, \bibnamefont{and}
  \bibinfo{author}{\bibfnamefont{K.}~\bibnamefont{Dransfeld}},
  \bibinfo{journal}{Phys. Lett.} \textbf{\bibinfo{volume}{42A}},
  \bibinfo{pages}{253} (\bibinfo{year}{1972}).

\bibitem[{\citenamefont{Laermans and Hunklinger}(1977)}]{Laermans1977}
\bibinfo{author}{\bibfnamefont{W.}~\bibnamefont{Laermans},
  \bibfnamefont{C.~Arnold}} \bibnamefont{and}
  \bibinfo{author}{\bibfnamefont{S.}~\bibnamefont{Hunklinger}},
  \bibinfo{journal}{J. Phys. C} \textbf{\bibinfo{volume}{10}},
  \bibinfo{pages}{L161} (\bibinfo{year}{1977}).

\bibitem[{\citenamefont{Golding and Graebner}(1976)}]{Golding1976}
\bibinfo{author}{\bibfnamefont{B.}~\bibnamefont{Golding}} \bibnamefont{and}
  \bibinfo{author}{\bibfnamefont{J.~E.} \bibnamefont{Graebner}},
  \bibinfo{journal}{Phys. Rev. Lett.} \textbf{\bibinfo{volume}{37}},
  \bibinfo{pages}{852} (\bibinfo{year}{1976}).

\bibitem[{\citenamefont{Armour and Blencowe}(2008)}]{Armour2008}
\bibinfo{author}{\bibfnamefont{A.~D.} \bibnamefont{Armour}} \bibnamefont{and}
  \bibinfo{author}{\bibfnamefont{M.~P.} \bibnamefont{Blencowe}},
  \bibinfo{journal}{New J. Phys.} \textbf{\bibinfo{volume}{10}},
  \bibinfo{pages}{095004} (\bibinfo{year}{2008}).

\end{thebibliography}
\end{document}